\documentclass[10pt, conference, letterpaper]{IEEEtran}
\IEEEoverridecommandlockouts

\def\BibTeX{{\rm B\kern-.05em{\sc i\kern-.025em b}\kern-.08em
    T\kern-.1667em\lower.7ex\hbox{E}\kern-.125emX}}
\ifCLASSINFOpdf
\fi
\usepackage{verbatim}
\usepackage{algorithm}
\usepackage{graphicx}
\usepackage{cite}
\usepackage{bm}
\usepackage{amssymb,amsfonts}
\usepackage{algorithmicx}
\usepackage[noend]{algpseudocode}
\usepackage{setspace}
\usepackage{float}
\usepackage{subfigure}
\usepackage{epstopdf}
\usepackage{amsthm}
\newtheorem{lemma}{Lemma}
\newtheorem{proposition}{Proposition}
\newtheorem{theorem}{Theorem}

\usepackage{stfloats}
\usepackage{enumerate}
\usepackage{multirow}
\usepackage{latexsym}
\usepackage{multicol,lipsum}
\usepackage{cuted} 

\usepackage{tikz}
\usepackage{textcomp}
\usepackage{xcolor}
\usepackage{cancel}

\pdfoutput=1
\usepackage[cmex10]{amsmath}
\usepackage[cmex10]{mathtools}

\algdef{SE}[DOWHILE]{Do}{doWhile}{\algorithmicdo}[1]{\algorithmicwhile\ #1}%
\allowdisplaybreaks
\makeatletter 
\def\@eqnnum{{\normalsize \normalcolor (\theequation)}} 

\makeatother 
\begin{document}
\title{Energy-Aware Vertically Layered Pinching-Antenna System Design with Probabilistic LoS Availability\vspace{-0.65em}}
\author{Ruotong Zhao and Derrick Wing Kwan Ng,~\IEEEmembership{Fellow,~IEEE} \\
\IEEEauthorblockA{School of Electrical Engineering and Telecommunications, University of New South Wales, Sydney, NSW 2052, Australia
\vspace{-7mm}}}

\maketitle

\begin{abstract}
This paper investigates a vertically layered pinching antenna (PA) system, in which dielectric waveguides deployed at different heights provide additional vertical spatial degrees of freedom for line-of-sight (LoS)-favorable PA placement. We consider a practical energy model that captures in-waveguide attenuation, base station-to-feed-point optical-fiber loss, and PA repositioning energy, along with a statistical LoS-availability model that depends on elevation and distance. We then jointly optimize the transmit beamforming, PA positions, time allocation, and transmission rates to maximize the expected successfully delivered frame-effective sum rate under per-user outage and quality-of-service constraints. To avoid enumerating the exponentially many LoS states, we derive a certified conditional-moment outage envelope for outage-guaranteed rate selection and a differentiable safe approximation of the chance constraints to facilitate tractable optimization. Building on these results, a block coordinate descent algorithm with a penalty-aided majorization-minimization procedure attains a high-quality suboptimal solution. Simulation results confirm the gains of vertical layering and reveal a fundamental energy-LoS tradeoff: LoS-favorable PA placement generally enhances both performance and reliability, although it incurs higher attenuation, path loss, and energy costs.

\end{abstract}

\section{Introduction}
Pinching antennas (PAs) have recently attracted significant attention for their spatial reconfigurability, which enables adaptive transmission in dynamic or obstructed scenarios within complex communication environments~\cite{zhao2026resource}.
Existing studies have investigated energy-constrained design for PA systems (PASS) in downlink~\cite{zeng2025energy} and uplink~\cite{zeng2025uplink} settings, demonstrating their advantages over conventional antenna architectures and characterizing the spectral--energy efficiency tradeoff~\cite{zhou2026spectral}.
In addition,~\cite{xu2025joint} minimized both the transmit and mechanical PA repositioning power. However, these energy-aware studies~\cite{zeng2025energy,zeng2025uplink,zhou2026spectral,xu2025joint} neglect position-dependent attenuation along the dielectric waveguide and the base station (BS)-to-feed-point optical-fiber loss. They also over-optimistically assume perfect line-of-sight (LoS) availability, ignoring inevitable blockage-induced randomness. As a result, they are unable to capture the fundamental non-trivial tradeoff among LoS-favorable PA placement, propagation conditions, and energy consumption.

On the other hand, recent studies have highlighted the potential of PASS to maintain LoS connectivity in blockage-prone environments~\cite{ding2025los,wang2025pin}, while largely neglecting PA repositioning energy costs. Although~\cite{li2026power} investigated a power-minimization problem under probabilistic LoS availability for PASS, it did not explicitly account for key practical energy consumption factors, e.g.,  motor-driven PA repositioning energy and the in-waveguide attenuation. In addition, these works predominantly adopt a conventional co-planar waveguide deployment, in which multiple waveguides are arranged in parallel at the same ceiling height~\cite{zeng2025energy,zeng2025uplink,zhou2026spectral,xu2025joint,blockage2025,ding2025los,wang2025pin,li2026power}, leaving the vertical deployment dimension unexploited. Specifically, higher PA elevations generally improve LoS availability by mitigating blockage, whereas lower elevations shorten the propagation distance and thereby reduce path loss~\cite{cai2022resource}. Moreover, moving a PA toward a LoS-favorable horizontal position can incur additional repositioning energy and in-waveguide attenuation.
This interplay among PA elevation, path loss, LoS availability, and energy consumption gives rise to a fundamental non-trivial resource allocation tradeoff that remains unexplored.

Motivated by the above observations, this work investigates a vertically layered multi-waveguide PASS under a practical energy model capturing in-waveguide attenuation, optical-fiber loss, and PA repositioning energy. We exploit statistical LoS availability for outage-guaranteed PASS design by jointly optimizing the transmit beamforming, PA positions, time allocation, and transmission rates to maximize the expected successfully delivered frame-effective sum rate, subject to maximum tolerable per-user outage probability constraints. 
Since exactly evaluating the outage probability of the formulated nonconvex chance-constrained problem requires enumerating exponentially many LoS states, we exploit the inherent link sparsity of PASS to derive a certified conditional-moment (CM) outage envelope for outage-guaranteed rate selection, and further construct a differentiable safe approximation of the chance constraints for tractable optimization. Capitalizing on these results, we develop an iterative block coordinate descent (BCD)-based algorithm incorporating a penalty-based method and majorization-minimization (MM), achieving a high-quality suboptimal solution. Numerical results show that the vertically layered PASS unlocks additional spatial degrees of freedom (DoF) and reveals a non-trivial tradeoff between energy consumption and LoS-favorable PA positioning under outage-guaranteed transmission.

\textbf{Notation:} 
Boldface capital and lowercase letters denote matrices and vectors, respectively. $\textbf{A}^\top$, $\textbf{A}^H$, $\text{Tr}(\textbf{A})$, and $\text{Rank}(\textbf{A})$ represent the transpose, Hermitian transpose, trace, and rank of $\textbf{A}$, respectively. 
The operator $\mathrm{diag}(\textbf{a})$ forms a diagonal matrix from vector $\textbf{a}$, while $\mathrm{Diag}(\textbf{A})$ extracts the main diagonal of matrix $\textbf{A}$.
$\mathbb{C}^{N \times M}$ ($\mathbb{R}^{N \times M}$),  $\mathbb{H}^{N}$, and $\mathbb{I}\{\cdot\}$ denote the spaces of $N \times M$ complex (real), $N \times N$ Hermitian matrices, and indicator function, respectively. The largest eigenvalue of $\textbf{A}$ and its associated eigenvector are $\lambda_{\rm max}\hspace{-1mm}\left(\hspace{-0.2mm}\textbf{A}\hspace{-0.2mm}\right)$ and $\bm{\lambda}_{\rm max}\hspace{-1mm}\left(\hspace{-0.2mm}\textbf{A}\hspace{-0.2mm}\right)$, respectively. We utilize $|\cdot|$, $\mathbb{E}\{\cdot\}$, $\textbf{1}$, $j = \sqrt{-1}$,  $\Re(a)$, and $\Im(a)$ to represent the absolute value, statistical expectation, the all-one vector, the imaginary unit, the real part, and the imaginary part of a complex number $a$, respectively. The circularly symmetric complex Gaussian (CSCG) distribution is denoted by $\mathcal{CN}(\mu, \sigma^2)$, with mean $\mu$ and variance $\sigma^2$.

\section{System Model and Problem Statement}\label{sec:System}
As illustrated in Fig. 1, we consider a downlink vertically layered PASS serving $K$ single-antenna users through $N$ dielectric waveguides mounted on a building surface, each supporting one movable PA. A central BS located at $\bm{\psi}^{\rm{BS}}=[0,0,z_{\rm BS}]^\top$ is connected via dedicated optical-fiber fronthaul links to phase-synchronized local feed modules colocated at the respective waveguide feed points. Unlike conventional co-planar PASS architectures~\cite{zeng2025energy,zeng2025uplink,zhou2026spectral,xu2025joint,blockage2025,ding2025los,wang2025pin,li2026power}, the waveguides are deployed at different heights, introducing an additional vertical dimension for effective PA placement. In a 3D Cartesian coordinate system, each waveguide has length $L$ meters, where the coordinates of the feed point and PA position on the $n$-th waveguide are $\bm{\psi}_n^{\mathrm{FP}} = [x_{\rm B},0,z_n]^\top$ and $\bm{\psi}_{n}^{\mathrm{Pin}} = [x^{\mathrm{Pin}}_{n},0,z_n]^\top$, respectively, with $z_n = z_1 + (n-1)d_z$ and $d_z > 0$ denoting the vertical separation between adjacent layers, $\forall n \in \mathcal{N} \triangleq \{1,\ldots,N\}$.
User $u_k$ is located at $\bm{\phi}_k = [x_k,y_k,0]^\top$, $\forall k \in \mathcal{K} \triangleq \{1,\ldots,K\}$. Let $s_k\in\mathbb{C}$, with $\mathbb{E}\{|s_k|^2\}=1$ and $\mathbb{E}\{s_j^{*}s_k\}=0$, $\forall j\neq k$, denote the information symbol for $u_k$ and let $\mathbf{w}_k=[w_{k,1},\ldots,w_{k,N}]^{\top}\in\mathbb{C}^{N\times 1}$ denote its beamforming vector, such that the transmit signal at the feed points is $\mathbf{s}\triangleq \sum_{k\in\mathcal{K}}\mathbf{w}_k s_k \in \mathbb{C}^{N\times 1}$.

\begin{figure}[t]
\centerline{\includegraphics[width=2in]{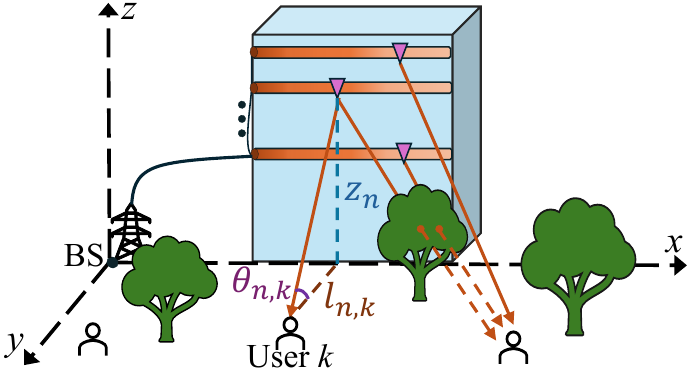}}
\vspace{-1em}
\caption{System architecture of the proposed vertically layered PASS.}
\vspace{-1.5em}
\label{fig:SYSTEM}
\end{figure}

\subsection{Hybrid LoS Availability Model}
For each PA-user link, the horizontal and 3D PA-user distances are $l_{n,k} = \sqrt{(x_{n}^{\mathrm{Pin}} - x_k)^2 + y_k^2}$ and $d_{n,k} = \sqrt{l_{n,k}^2 + z_n^2}$, respectively, with the corresponding elevation angle $\theta_{n,k} = \tfrac{180}{\pi}\tan^{-1}(\tfrac{z_n}{l_{n,k}})$, $\forall n, k$, and $\theta_{n,k} = 90^{\circ}$ when $l_{n,k} = 0$. The LoS availability of each link is modeled by the multiplicative hybrid probability $q_{n,k}^{\mathrm{LoS}} = q_{n,k}^{\mathrm{ele}}q_{n,k}^{\mathrm{dis}}$ since $q_{n,k}^{\mathrm{LoS}} \le \min\{q_{n,k}^{\mathrm{ele}}, q_{n,k}^{\mathrm{dis}}\}$, where $q_{n,k}^{\mathrm{ele}} = \tfrac{1}{1+A_{\rm L}\exp[-B_{\rm L}(\theta_{n,k}-A_{\rm L})]}$~\cite{cai2022resource, bor2016efficient} and $q_{n,k}^{\mathrm{dis}} = \exp(-\wp_{\rm L}d_{n,k}^{2})$~\cite{li2026power} capture elevation-dependent LoS visibility and distance-dependent blockage, respectively, with LoS parameters $A_{\rm L}, B_{\rm L} \ge 0$ and blockage density $\wp_{\rm L} \ge 0$. The LoS indicators $\{\chi_{n,k}\}_{n,k}$ are modeled as independent Bernoulli variables with $\Pr\{\chi_{n,k}=1\}=q_{n,k}^{\mathrm{LoS}}$~\cite{li2026power}. Collecting $\bm{\chi}_k = [\chi_{1,k}, \ldots, \chi_{N,k}]^\top$, $\bm{\chi} \triangleq \{\bm{\chi}_{k}\}_{k \in \mathcal{K}}$, each realization occurs, conditioned on the PA-position vector $\textbf{x}^{\rm Pin}$, with probability $\pi_k(\bm{\chi}_k|\mathbf x^{\rm Pin},\bm\phi_k) =  \underset{{n \in \mathcal{N}}}{\Pi} (q_{n,k}^{\rm LoS})^{\chi_{n, k}} (1-q_{n,k}^{\rm LoS})^{1-\chi_{n, k}}$. Unlike conventional distance-based PASS blockage models~\cite{li2026power}, this hybrid model incorporates elevation-angle dependence, thereby capturing the dependence of the probabilistic quality-of-service (QoS) constraints on PA positioning.

\subsection{Energy Consumption and Power-Delivery Model}
\begin{figure}[t]
\centerline{\includegraphics[width=2.2in]{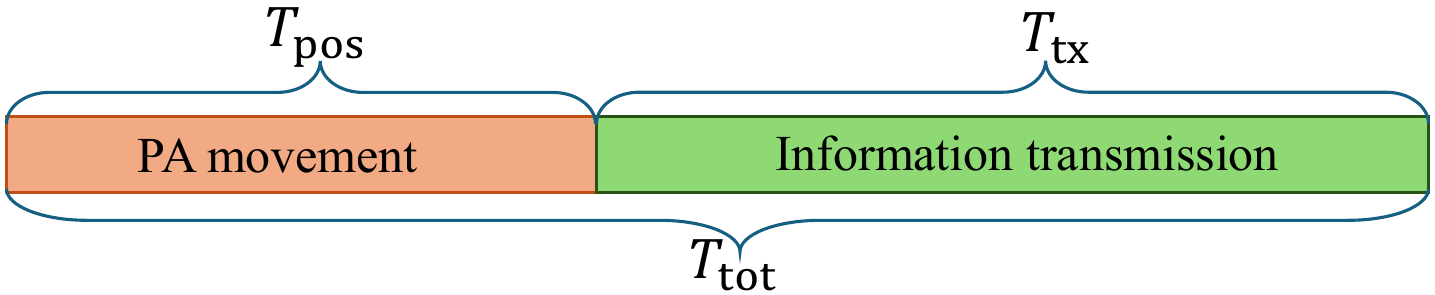}}
\vspace{-1em}
\caption{Illustration of the proposed two-phase protocol for PA movement
and information transmission.}
\vspace{-1.5em}
\label{fig:Time}
\end{figure}
As shown in Fig.~2, we adopt a two-phase protocol~\cite{xu2025joint} with PA positioning, information transmission, and total frame durations $T_{\rm pos}$, $T_{\rm tx}$, and $T_{\rm tot} \ge T_{\rm pos} + T_{\rm tx}$, respectively. The position-dependent power attenuation of the $n$-th PA is $p_{n} = e^{-2\alpha\|\bm{\psi}_n^{\mathrm{FP}}-\bm{\psi}_{n}^{\mathrm{Pin}}\|_2}$~\cite{hu2025}, where $\alpha = \lambda_g \epsilon_r \pi f_c^2c^{-2} \text{tan}(\delta_e)\in \mathbb{R}^+$ is the attenuation constant, with dielectric constant $\epsilon_r$ and effective dielectric loss tangent $\text{tan}(\delta_e)$. With previous PA position $\tilde{\bm{\psi}}_{n}^{\mathrm{Pin}}=[\tilde{x}_{n}^{\mathrm{Pin}},0,z_n]^\top$ and displacement $\Delta_{n}= |x_n^{\rm Pin} - \tilde{x}_n^{\rm Pin}|$, the circuit and PA movement energies are $E_{\rm Cir} = N(T_{\rm tx} + T_{\rm pos})P_{\rm cir}$ and $E_{\rm Mov} = \frac{P_{\mathrm{mot}}\sum_{n=1}^{N}\Delta_{n}}{v_{\mathrm{PA}}}$~\cite{xu2025joint}, respectively, where $P_{\rm cir}$, $P_{\rm mot} > 0$, and $v_{\rm PA} > 0$ denote the static circuit power, per-PA motor power, and constant moving speed, respectively. The $n$-th fronthaul link has length and power-delivery factor $\ell_n^{\rm{fib}}=\left\|\bm{\psi}_n^{\mathrm{FP}}-\bm{\psi}^{\rm{BS}}\right\|_2$ and $\eta_n^{\rm{fib}}=e^{-2\alpha_{\mathrm{f}}\ell_n^{\rm{fib}}}$, respectively, where $\alpha_{\rm{f}} =\frac{\ln 10}{20}\kappa_{\rm{f}}$ with fiber attenuation coefficient $\kappa_{\rm{f}}$. Hence, the total consumed energy is
\begin{align}
E_{\mathrm{total}} = \underbrace{\frac{T_{\mathrm{tx}}}{\eta_{\mathrm{BS}}} \sum_{n\in\mathcal{N}} \frac{1}{\eta^{\rm{fib}}_n} \sum_{k\in\mathcal{K}} |w_{k,n}|^2 }_{\text{Transmission},\ E_{\mathrm{Tx}}} +E_{\mathrm{Cir}}+E_{\mathrm{Mov}},
\end{align}
where $\eta_{\mathrm{BS}}\in(0,1]$ is the power-amplifier efficiency at the BS. Importantly, the LoS-favorable PA positions are generally not aligned with the energy-minimizing ones, and this tradeoff jointly shapes the subsequent resource allocation.

\subsection{Channel and Signal Model}
We assume that the BS knows the user locations and the statistical LoS availability of all PA--user links~\cite{ding2025los}. For the $n$-th PA--$u_k$ link, the downlink spherical-wave near-field channel and in-waveguide phase-shift coefficient are~\cite{blockage2025}
\begin{align}\label{eq:channel_user}
h_{n,k}^{\rm L} = \tfrac{ \eta^{1/2} e^{-j\frac{2\pi}{\lambda}\|\bm{\phi}_k-\bm{\psi}_{n}^{\rm Pin}\|}}
     {\|\bm{\phi}_k-\bm{\psi}_{n}^{\rm Pin}\|}, \quad
g_{n} \hspace{-1mm}=\hspace{-0.8mm} e^{-j\frac{2\pi}{\lambda_g}\hspace{-0.5mm}\left\|\hspace{-0.4mm}\bm{\psi}_n^{\rm FP} \hspace{-0.8mm}-\hspace{-0.2mm} \bm{\psi}_{n}^{\rm Pin} \hspace{-0.6mm}\right\|}, \hspace{0.2mm}
\end{align}
respectively, where $\eta=\tfrac{c^2}{16\pi^2f_c^2}$ is the free-space reference channel gain, $\lambda_g=\lambda/\eta_{\mathrm{eff}}$, and $c$, $f_c$, $\lambda$, and $\eta_{\mathrm{eff}}>1$ denote the speed of light, carrier frequency, wavelength, and effective refractive index, respectively~\cite{hu2025}. The blockage-aware channel $h_{n,k}(\chi_{n,k})=\chi_{n,k}h_{n,k}^{\mathrm{L}} \in \mathbb{C}$ models an LoS-dominant regime in which blocked links have negligible non-LoS components~\cite{zhao2026resource}. Accordingly, the equivalent downlink channel of $u_k$ is $\bm{\beta}_k (\bm{\chi}_{k}) = [\beta_{k,1}(\chi_{1,k}),\ldots,\beta_{k,N}(\chi_{N,k})]^\top  \in \mathbb{C}^{N \times 1}$ with $\beta_{k,n}(\chi_{n,k}) = (e^{-\alpha\|\bm{\psi}_n^{\mathrm{FP}}-\bm{\psi}_{n}^{\mathrm{Pin}}\|_2} g_{n}h_{n,k}(\chi_{n,k}) )^*$, $\forall n,k$, and the received signal at $u_k$ is
\begin{align}\label{eq:rx_sig}
\varrho_k = \bm{\beta}_{k}^{H} (\bm{\chi}_{k}) \mathbf{s} + \epsilon_k = \sum_{j\in\mathcal{K}}\bm{\beta}_{k}^{H} (\bm{\chi}_{k})\mathbf{w}_j s_j + \epsilon_k, \forall k,
\end{align}
where $\epsilon_k\sim\mathcal{CN}(0,\sigma_k^2)$ is the additive white Gaussian noise with power $\sigma_k^2$.

\subsection{Problem Formulation}
For a given LoS realization $\bm{\chi}$, the achievable rate of $u_k$ in bit/s/Hz is $R_k(\bm{\chi}_{k}) = \log_2 (1 + \gamma_k(\bm{\chi}_{k}))$ with the SINR
\begin{align}
\gamma_k(\bm{\chi}_{k}) = \tfrac{|\bm{\beta}_k^H (\bm{\chi}_{k}) \mathbf{w}_k|^2}{\sum_{j \neq k} |\bm{\beta}_k^H (\bm{\chi}_{k}) \mathbf{w}_j |^2+ \sigma_k^2}, \forall k.
\end{align}
Let $r_k$ denote the frame-effective transmission rate of $u_k$, corresponding to a coding rate $\tfrac{T_{\rm tot}}{T_{\rm tx}}r_k$ during $T_{\rm tx}$, such that its transmission is successful only if $\tfrac{T_{\rm tx}}{T_{\rm tot}}R_k(\bm{\chi}_k)\ge r_k$. We thus adopt the expected successfully delivered frame-effective sum rate as the performance metric, averaged over $\bm{\chi}$ conditioned on $\mathbf{x}^{\rm Pin} \hspace{-1mm}=\hspace{-1mm}[x_1^{\rm Pin}\hspace{-1mm},\ldots,x_N^{\rm Pin}]^{\hspace{-0.8mm}\top}\hspace{-1mm}$, capturing PA repositioning time, selected transmission rates, and LoS randomness, i.e.,

\begin{align}
\bar{R}  =   \mathbb{E}_{\bm{\chi} | \mathbf{x}^{\rm Pin}}  \Big[ \sum\nolimits_{k \in \mathcal{K}}  r_k \mathbb{I}\left\{\tfrac{T_{\rm tx}}{T_{\rm tot}} R_k(\bm{\chi}_{k}) \ge r_k \right\} \Big].
\end{align}
Accordingly, the joint beamforming, PA-positioning, time-allocation, and transmission-rate design is formulated as
\begin{align}
&\underset{\textbf{w}_k, \ r_k, \ x^{\rm{Pin}}_{n},\ T_{\rm pos}, \ T_{\rm tx}} {\rm{maximize}} \hspace{2mm} \bar{R}  \nonumber \\
\rm{s.t.}\ &({\rm C1})\hspace{-1mm}:\hspace{-0.7mm} x_{\rm B} \hspace{-1mm}\leq \hspace{-0.8mm} x_{n}^{\rm{Pin}} \hspace{-0.9mm} \leq \hspace{-1mm} x_{\rm B} \hspace{-1mm}+ \hspace{-1mm}L, \hspace{-0.8mm}\forall \hspace{-0.3mm} n\hspace{-0.3mm}, 
({\rm C2})\hspace{-1.2mm}: \hspace{-1.1mm}| \hspace{-0.3mm} x_{n}^{\rm{Pin}} \hspace{-1.2mm}-\hspace{-0.8mm} \Tilde{x}_{n}^{\rm{Pin}} \hspace{-0.4mm}| \hspace{-1mm}\le\hspace{-1.1mm} v_{\rm PA} \hspace{-0.5mm} T_{\hspace{-0.7mm} {\rm pos}}\hspace{-0.4mm}, \hspace{-0.7mm}\forall \hspace{-0.3mm}n\hspace{-0.2mm}, \nonumber \\
&({\rm C3})\hspace{-1mm}:\hspace{-0.7mm} \sum_{n \in \mathcal{N}} \frac{1}{\eta^{\rm fib}_n} \sum_{k \in \mathcal{K}} |w_{k,n}|^2 \leq\hspace{-0.5mm} P^{\max}_{\rm Tx}\hspace{-0.5mm}, \nonumber \\
&({\rm C4})\hspace{-1mm}:\hspace{-0.7mm} \ E_{\rm total} \leq\hspace{-0.5mm} E^{\max}_{\rm total}, \ ({\rm C5})\hspace{-1mm}:\hspace{-0.7mm} T_{\rm pos} + T_{\rm tx} \le T_{\rm tot}, \nonumber \\
&({\rm C6})\hspace{-1mm}:\hspace{-0.5mm} 
\underset{\bm{\chi}_k| \textbf{x}^{\rm Pin}, \bm{\phi}_k}{\text{Pr}} 
\left\{\frac{T_{\rm tx}}{T_{\rm tot}} R_k(\bm{\chi}_{k}) \ge r_k \right\} 
\ge 1 - \delta_k, \, \forall k, \nonumber \\
&({\rm C7})\hspace{-1mm}:\hspace{-0.5mm} 
r_k \ge R_{\min,k}^{\mathrm{Req}}, \, \forall k.
\label{eq: formulation}
\end{align}
Constraints (C1) and (C2) confine each PA to its waveguide and limit its displacement to the distance reachable within $T_{\rm pos}$. Constraint (C3) limits the fronthaul-aware BS transmit power to $P_{\rm Tx}^{\rm max}$. Constraints (C4) and (C5) impose the total frame-energy budget $E_{\rm total}^{\rm max}$ and the time allocation. Constraint (C6) guarantees that the selected frame-effective rate $r_k$ is supported by the instantaneous blockage-aware channel with probability at least $1-\delta_k$; while constraint (C7) enforces the per-user minimum rate $R_{\min,k}^{\mathrm{Req}}$.
\begin{figure}[t]
\centerline{\includegraphics[width=3.1in]{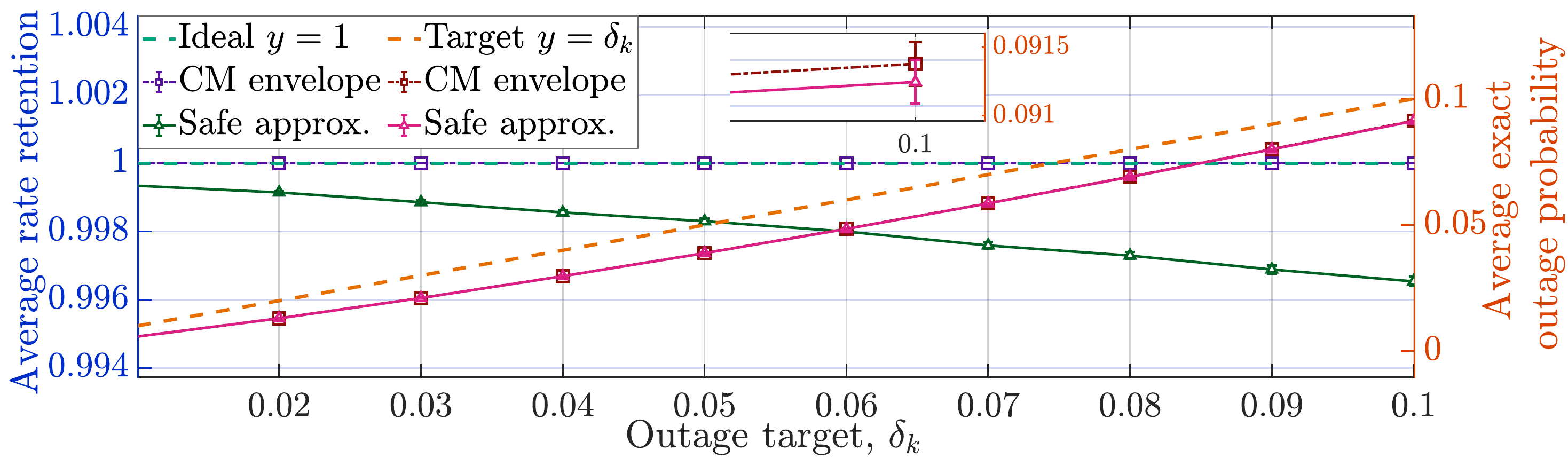}}
\vspace{-2mm}
\caption{Certified-outage verification of the CM envelope and the smooth safe approximation versus the outage target $\delta_k$, averaged over 10,000 random user-location trials under $N = 8$ and $K = 3$ in an urban setting~\cite{bor2016efficient,li2026power}.}
\vspace{-1.8em}
\label{fig:cm_delta}
\end{figure}

\section{Problem Solution}
Problem~\eqref{eq: formulation} is highly nonconvex due to the strongly coupled variables. In particular, since exact evaluation of the outage probability in (C6) requires enumerating all $2^N$ LoS states per user, we develop a CM outage envelope, i.e., a certified upper bound, for outage-guaranteed rate selection, and a differentiable safe approximation of (C6) for tractable optimization.

\subsection{Outage-Guaranteed Rate Selection}
\label{subsec:outage_rate_selection}
For $T_{\rm tx}>0$ and $0\le\delta_k<1$, define the coding rate $\rho_k\triangleq\frac{T_{\rm tot}}{T_{\rm tx}}r_k$ and the outage probability $F_k(\rho^-)\triangleq\Pr_{\bm\chi_k|\mathbf x^{\rm Pin},\bm\phi_k}\{R_k(\bm\chi_k)<\rho\}$ where $\rho\ge0$, with the dependence on $(\mathbf W,\mathbf x^{\rm Pin})$ suppressed. Then, (C6) is equivalent to $F_k(\rho_k^-)\le\delta_k$, $\forall k$, imposing for fixed $(\mathbf W,\mathbf x^{\rm Pin})$ the exact rate ceiling $\rho_k^{\rm C6}\triangleq\max\{\rho\ge0:F_k(\rho^-)\le\delta_k\}$. To avoid direct enumeration of all $2^N$ LoS states during rate selection, motivated by the dominant-link structure of PASS~\cite{blockage2025} and the principle of scenario reduction~\cite{dupavcova2003scenario}, we group the LoS states into sparse subsets and classify each subset with a certified rate interval as successful, in outage, or ambiguous.
\begin{theorem}\label{TH: 1}
For fixed $(\textbf{W}_k,\textbf{x}^{\rm Pin})$, let $\mathcal I_k^{\rm sp}$ be a partition of $\{0,1\}^N$, where subset $i$ occurs with probability $p_{i,k}(\mathbf x^{\rm Pin})$, fixes $\chi_{n,k}=b_{i,n}\in\{0,1\}$ for $n\in\mathcal A_i$, and leaves $\mathcal R_i=\mathcal N\setminus\mathcal A_i$ as independent Bernoulli variables. If the signal and interference-plus-noise powers lie in $[S_{i,k}^{\rm L},S_{i,k}^{\rm U}]$ and $[I_{i,k}^{\rm L},I_{i,k}^{\rm U}]$, respectively, then subset $i$ admits the certified rate interval
\begin{align}
R_{i,k}^{\rm L} \hspace{-1.3mm}=\hspace{-1mm} \log_2 (\hspace{-0.5mm}1 \hspace{-1mm}+\hspace{-0.8mm} \tfrac{S_{i,k}^{\rm L}}{I_{i,k}^{\rm U}} \hspace{-0.6mm}) \hspace{-1mm}\le\hspace{-1mm} R_k(\bm{\chi}_k) \hspace{-1mm}\le\hspace{-1mm}
\log_2(1 \hspace{-1mm}+\hspace{-1mm} \tfrac{S_{i,k}^{\rm U}}{I_{i,k}^{\rm L}}) \hspace{-1mm}=\hspace{-1mm} R_{i,k}^{\rm U}.
\label{eq:subset_rate_bounds}
\end{align}
Let $\mathcal U_{i,k}\triangleq(R_{i,k}^{\rm L},R_{i,k}^{\rm U}]$ and let the margin $Z_k(\bm\chi_k;\rho)=|\bm\beta_k^H(\bm\chi_k)\mathbf w_k|^2-(2^{\rho}-1)\big(\sum_{j\ne k}|\bm\beta_k^H(\bm\chi_k)\mathbf w_j|^2+\sigma_k^2\big)$, satisfying $R_k(\bm\chi_k)<\rho\Leftrightarrow Z_k<0$, have conditional mean $\mu_i(\rho)$ and variance $v_i(\rho)$ in subset $i$. With the CM contribution
\begin{align}
\psi_{i,k}^{\rm CM}(\rho)\hspace{-1mm}=\hspace{-1mm}
\begin{cases}
0, & \rho\le R_{i,k}^{\rm L},\\[0.3mm]
\psi_{i,k}^{\rm ex}(\rho), & \rho\in\mathcal U_{i,k}, |\mathcal R_i|\le m_{\rm ex},\\[0.3mm]
\tfrac{v_i(\rho)}{v_i(\rho)+\mu_i^2(\rho)}, & \rho\in\mathcal U_{i,k}, |\mathcal R_i|>m_{\rm ex}, \mu_i>0,\\[0.3mm]
1, & \text{otherwise},
\end{cases}\label{eq:psiCM}
\end{align}
Then, the CM envelope $\bar F_k^{\rm CM}(\rho)\triangleq\sum_{i}p_{i,k}\psi_{i,k}^{\rm CM}(\rho)$ satisfies $F_k(\rho^-)\le\bar F_k^{\rm CM}(\rho)$, such that any $\rho_k^\star$ with $\bar F_k^{\rm CM}(\rho_k^\star)\le\delta_k$ guarantees (C6) with $r_k^\star=\tfrac{T_{\rm tx}}{T_{\rm tot}}\rho_k^\star$.
\end{theorem}
\vspace{-1em}
\begin{proof}
With $\beta_{k,n}(\chi_{n,k}) = \chi_{n,k}\bar{\beta}_{n,k}$ and $a_{n,j,k}=\bar\beta_{n,k}^{*}w_{j,n}$, each stream amplitude in subset $i$ lies in the rectangle obtained by adding $\sum_{n\in\mathcal A_i}b_{i,n}a_{n,j,k}$ to the componentwise real/imaginary interval sums of $\{a_{n,j,k}\}_{n \in \mathcal{R}_i}$, whose minimum and maximum squared distances to the origin yield $S_{i,k}^{\rm L/U}$ and $I_{i,k}^{\rm L/U}$, establishing \eqref{eq:subset_rate_bounds}. Hence, the conditional outage is $0$ for $\rho\le R_{i,k}^{\rm L}$, $1$ for $\rho>R_{i,k}^{\rm U}$, and, for $\rho\in\mathcal U_{i,k}$, evaluated exactly or bounded via Cantelli's inequality~\cite{ghosh2002probability}; summing over the partition completes the proof.
\end{proof}
\vspace{-0.3em}
As the Cantelli term may render $\bar F_k^{\rm CM}$ non-monotone in $\rho$, over a finite set $\mathcal A_k^{\rm cand}$ collecting $0$ and the boundaries in \eqref{eq:subset_rate_bounds}, we adopt a backward-min regularization 
$\bar F_k^{\downarrow}(\rho)=\min_{\rho'\ge\rho}\bar F_k^{\rm CM}(\rho')$, 
such that $F_k(\rho^-)\le\bar F_k^{\rm CM}(\rho')$, $\forall\rho'\ge\rho$, and select $\rho_k^{\rm CM}=\max\{\rho\in\mathcal A_k^{\rm cand}:\bar F_k^{\downarrow}(\rho)\le\delta_k\}$. Being non-smooth, $\bar F_k^{\rm CM}$ is unsuited to gradient-based optimization; we thus replace the outage indicator by the smooth upper bound $\phi_{\tau,\kappa}(s)=\frac{{\rm sig}(\kappa-s/\tau)}{{\rm sig}(\kappa)}\ge\mathbb I\{s<0\}$, with ${\rm sig}(x)=\frac{1}{1+e^{-x}}$~\cite{blockage2025} and smoothing parameters $\tau,\kappa>0$.
\vspace{-0.3em}

\begin{proposition}\label{pro: 1}
Let $\mathcal M_k$ be a partition of $\{0,1\}^N$, where $\ell\in\mathcal M_k$ has probability $p_{\ell,k}(\mathbf x^{\rm Pin})$ with $R_{\ell,k}^{\rm sp}\le R_k(\bm\chi_k)$ for all $\bm\chi_k$ represented by $\ell$. Then, the safe outage approximation
\begin{align}
\widetilde F_k^{\rm SA}(\rho;\hspace{-1mm}\mathbf W\hspace{-1mm},\mathbf x^{\rm Pin}) \hspace{-1mm}=\hspace{-3mm} \sum_{\ell\in\mathcal M_k} \hspace{-1.7mm} p_{\ell,k}(\hspace{-0.5mm}\mathbf x^{\rm Pin}\hspace{-0.5mm}) \phi_{\tau,\kappa} \hspace{-1.4mm}\left(\hspace{-0.7mm} R_{\ell,k}^{\rm sp}(\mathbf W,\mathbf x^{\rm Pin})\hspace{-0.7mm}-\hspace{-0.7mm}\rho \hspace{-0.7mm}\right)
\label{eq:soft_CM_envelope}
\end{align}
satisfies $F_k(\rho^-)\le\widetilde F_k^{\rm SA}(\rho;\mathbf W,\mathbf x^{\rm Pin})$, so that $(\widetilde{\rm C6})\hspace{-1mm}:\hspace{-0.5mm} \widetilde F_k^{\rm SA}(\rho_k;\mathbf W,\mathbf x^{\rm Pin})\le\delta_k$, $\forall k$, is a safe approximation of (C6).
\end{proposition}
\vspace{-1.3em}
\begin{proof}
Since $R_{\ell,k}^{\rm sp}\le R_k(\bm\chi_k)$ for every LoS state and $\phi_{\tau,\kappa}(s)\ge\mathbb I\{s<0\}$, we have $\mathbb I\{R_k(\bm\chi_k)<\rho\}\le\phi_{\tau,\kappa}(R_{\ell,k}^{\rm sp}-\rho)$; averaging over the partition then proves the result.
\end{proof}

\textbf{Remark 1.}
Theorem 1 yields a certified CM envelope for rate selection and outage verification, whereas the full-atom instance of Proposition 1 yields a differentiable safe surrogate for resource allocation while avoiding sorting at the cost of $2^N$ evaluations per user. Fig.~\ref{fig:cm_delta} shows that $\rho_k^{\rm CM}$ nearly coincides with the exact benchmark $\rho_k^{\rm C6}$, while the safe-approximation rate remains close, and the correspondibng outages of both stay below $\delta_k$. Since near-field channels exhibit the same domination by a few effective components~\cite{cui2022channel}, the proposed framework generalizes beyond the considered PASS.

Enabled by Proposition~\ref{pro: 1}, we henceforth replace the intractable chance constraint (C6) in~\eqref{eq: formulation} by its safe approximation $(\widetilde{\rm C6})$, whose satisfaction guarantees (C6), and directly optimize $\bm\rho\triangleq[\rho_1,\ldots,\rho_K]^{\rm T}$ with $r_k=\frac{T_{\rm tx}}{T_{\rm tot}}\rho_k$. 
We then lift $\mathbf w_k$ into $\mathbf W_k\triangleq\mathbf w_k\mathbf w_k^{H}$ with $\mathcal W\triangleq\{\mathbf W_k\}_{k\in\mathcal K}$ and $[\mathbf W_k]_{n,n}=|w_{k,n}|^2$. For a fully refined partition $\mathcal M_k^{\rm full}$, whose atom rate is $R_{\ell,k}^{\rm sp}(\mathcal W,\mathbf x^{\rm Pin})=\log_2\!\big(1+\frac{{\rm Tr}(\mathbf B_{\ell,k}\mathbf W_k)}{\sum_{j\ne k}{\rm Tr}(\mathbf B_{\ell,k}\mathbf W_j)+\sigma_k^2}\big)$ with $\mathbf B_{\ell,k}(\mathbf x^{\rm Pin})\triangleq\mathbf D_\ell\bar{\bm\beta}_k\bar{\bm\beta}_k^{H}\mathbf D_\ell$, where $\mathbf D_\ell = \text{diag}(b_{\ell,1}, ..., b_{\ell,N})$ is the diagonal LoS-pattern matrix of atom $\ell$ and $\bar{\bm\beta}_k\triangleq[\bar\beta_{1,k},\ldots,\bar\beta_{N,k}]^{\rm T}$. Since $\phi_{\tau,\kappa}(\cdot)$ is nonincreasing, $(\widetilde{\rm C6})$ can be equivalently reformulated by $(\widetilde{\rm C6a})$--$(\widetilde{\rm C6b})$ in \eqref{eq: formulation1} with slack variables $\bm\varepsilon\triangleq\{\varepsilon_{\ell,k}\in\mathbb R\}$, as $(\widetilde{\rm C6a})$ implies $\phi_{\tau,\kappa}(R_{\ell,k}^{\rm sp}-\rho_k)\le\phi_{\tau,\kappa}(\varepsilon_{\ell,k})$ while equality $\varepsilon_{\ell,k}=R_{\ell,k}^{\rm sp}-\rho_k$ recovers $(\widetilde{\rm C6})$. Moreover, since $(\widetilde{\rm C6})$ implies $F_k(\rho_k^-)\le\delta_k$, (5) yields $\bar R\ge\sum_{k\in\mathcal K}(1-\delta_k)\frac{T_{\rm tx}}{T_{\rm tot}}\rho_k$, leading to the conservative design
\begin{align}
&\underset{\mathcal{W}, \ \bm{\rho}, \ \mathbf{x}^{\rm{Pin}},\ T_{\rm pos}, \ T_{\rm tx}, \ \bm{\varepsilon}} {{\rm maximize}} \hspace{2mm} \sum\nolimits_{k \in \mathcal{K}}(1-\delta_k)\tfrac{T_{\rm tx}}{T_{\rm tot}}\rho_k  \nonumber \\
&{\rm s.t.}\ ({\rm C1})-({\rm C5}), (\widetilde{\rm C6a})\hspace{-1mm}:\hspace{-0.5mm} R_{\ell,k}^{\rm sp}(\mathcal W,\mathbf x^{\rm Pin})-\rho_k \ge \varepsilon_{\ell,k}, \ \forall \ell, k, \nonumber \\
&(\widetilde{\rm C6b})\hspace{-1mm}:\hspace{-3.5mm} \sum_{\ell\in\mathcal M_k} \hspace{-2.1mm} p_{\hspace{-0.4mm}\ell\hspace{-0.2mm},k} \hspace{-0.4mm}(\hspace{-0.4mm}\mathbf x^{\rm Pin}\hspace{-0.4mm}) \phi_{\tau,\kappa} \hspace{-1mm}\left(\hspace{-0.4mm}\varepsilon_{\ell, k}\hspace{-0.4mm}\right) \hspace{-1mm}\le\hspace{-1mm} \delta_k\hspace{-0.3mm},\hspace{-0.8mm} \forall k\hspace{-0.3mm}, \hspace{-0.4mm}(\widetilde{\rm C7})\hspace{-1.2mm}:\hspace{-1mm} \tfrac{T_{\hspace{-0.4mm}{\rm tot}}\hspace{-0.3mm}R_{\min,k}^{\rm Req}}{T_{\rm tx}} \hspace{-0.8mm}-\hspace{-1mm} \rho_k \hspace{-1mm}\le\hspace{-1mm} 0,\hspace{-1mm} \forall k\hspace{-0.4mm}, \nonumber \\
&({\rm C8})\hspace{-1mm}:\hspace{-0.5mm} \mathbf{W}_k\succeq \bm{0}, \ \forall k, \ ({\rm C9})\hspace{-1mm}:\hspace{-0.5mm} {\rm Rank}(\mathbf{W}_k) \le 1, \ \forall k.
\label{eq: formulation1}
\end{align}
As such, any feasible solution to \eqref{eq: formulation1} is also feasible for \eqref{eq: formulation}.
Since \eqref{eq: formulation1} remains nonconvex, we adopt the BCD method to alternately update $\{\bm\rho,\mathcal W,T_{\rm tx},T_{\rm pos},\bm\varepsilon\}$ in subproblem 1 and $\{\bm{\rho}, \mathbf{x}^{\rm Pin}, \bm{\varepsilon}, T_{\rm pos}, T_{\rm tx}\}$ in subproblem 2.

\subsection{Solution of $\{\bm{\rho}, \mathcal{W}, T_{\rm tx}, T_{\rm pos}, \bm{\varepsilon}\}$}
\label{subsec:subprob1}
Even after BCD, subproblem 1 remains nonconvex owing to the bilinear coupling in the objective and $({\rm C4})$, the difference-of-convex (DC) structure of $(\widetilde{\rm C6a})$, the sigmoid-sum constraint $(\widetilde{\rm C6b})$, and the rank-one constraint $({\rm C9})$. Since $T_{\rm tx}>0$, we introduce $\textbf{Q}_k = T_{\rm tx} \textbf{W}_k$, $\forall k$, with $\mathcal{Q} \triangleq \{\mathbf Q_1, \ldots, \mathbf Q_K\}$, which renders $({\rm C3})$ and $({\rm C4})$ affine and recasts subproblem 1 as
\begin{align}\label{eq: formulation12}
&\underset{\mathcal{Q}, \ \bm{\rho}, \ T_{\rm pos}, \ T_{\rm tx}, \ \bm{\varepsilon}} {\rm{maximize}} \sum\nolimits_{k\in\mathcal K}\tfrac{ (1-\delta_k)}{4 T_{\rm tot}} \left[ (T_{\rm tx}+\rho_k)^2 - (T_{\rm tx}-\rho_k)^2\right] \nonumber \\
&\rm{s.t.}\ ({\rm C2}), ({\rm C5}), (\widetilde{\rm C6b}), (\widetilde{\rm C7}), \nonumber \\
&(\widetilde{\rm C3})\hspace{-1mm}:\hspace{-0.7mm} \sum_{n \in \mathcal{N}} \frac{1}{\eta^{\rm fib}_n} \sum_{k \in \mathcal{K}} [\textbf{Q}_k]_{n,n} \leq\hspace{-0.5mm} P^{\max}_{\rm Tx} T_{\rm tx}\hspace{-0.5mm}, \nonumber \\
&(\widetilde{\rm C4})\hspace{-1mm}:\hspace{-0.7mm} \frac{1}{\eta_{\mathrm{BS}}} \sum_{n \in \mathcal{N}} \frac{1}{\eta^{\rm fib}_n} \sum_{k \in \mathcal{K}} [\textbf{Q}_k]_{n,n} \hspace{-1mm}+\hspace{-1mm} {E}_{\rm Cir} \hspace{-1mm}+\hspace{-1mm} {E}_{\rm Mov} \leq\hspace{-0.5mm} E^{\max}_{\rm total} , \nonumber \\
&(\widetilde{\rm C6a})\hspace{-1mm}:\hspace{-0.5mm} \log_2 \Big(\sum\nolimits_{i \in \mathcal{K}}{\rm Tr}(\mathbf B_{\ell,k}\mathbf Q_i)+ T_{\rm tx}\sigma_k^2 \Big) - \rho_k \nonumber \\
& \hspace{8mm} -\log_2 \hspace{-1mm} \Big(\hspace{-1mm}\sum\nolimits_{j\ne k}\hspace{-1mm}{\rm Tr}(\mathbf B_{\ell,k}\mathbf Q_j) \hspace{-1mm}+\hspace{-1mm} T_{\rm tx}\sigma_k^2\Big) \hspace{-1.3mm}\ge\hspace{-1mm} \varepsilon_{\ell,k}, \forall \ell, k, \nonumber \\
&({\rm C8})\hspace{-1mm}:\hspace{-0.5mm} \textbf{Q}_k\succeq \bm{0}, \forall k, \ ({\rm C9})\hspace{-1mm}:\hspace{-0.5mm} {\rm Rank}( \textbf{Q}_k) \le 1, \forall k.
\end{align}
To handle constraint $({\rm C9})$, we invoke the following lemma.
\vspace{-0.5em}
\begin{lemma}\label{lem:rankone}
The rank-one constraint $({\rm C9})$ is equivalent to the DC constraint $(\widetilde{{\rm C9}}): \ \left\|\textbf{Q}_k\right\|_* - \left\|\textbf{Q}_k\right\|_2 \leq 0, \ \forall k$.
\end{lemma}
\vspace{-1em}
\begin{proof}
For any $\textbf{X} \in \mathbb{H}^n$ with $\textbf{X}\succeq \textbf{0}$, $\left\|\textbf{X}\right\|_* = \sum_i x_i \geq \left\|\textbf{X}\right\|_2 = \max_i\{x_i\}$, where $x_i$ is the $i$-th singular value of $\textbf{X}$, with equality if and only if $\text{Rank}(\textbf{X}) \le 1$.
\end{proof}

\vspace{-1em}
We convexify subproblem 1 via the MM framework~\cite{blockage2025}, for which we first establish a reusable surrogate result.
\vspace{-0.5em}
\begin{lemma}\label{lem:mm}
Let $\mathbf y^{(r)}$ be the $r$-th MM iterate, $\langle \mathbf A, \mathbf B\rangle \triangleq \Re\{{\rm Tr}(\mathbf A^H \mathbf B)\}$, and $\mathcal L_h^{(r)}(\mathbf y) \triangleq h(\mathbf y^{(r)}) + \langle \nabla h(\mathbf y^{(r)}), \mathbf y - \mathbf y^{(r)}\rangle$ be the first-order expansion of differentiable $h$. Then, \emph{(i)} $f(\mathbf y) \ge \mathcal L_f^{(r)}(\mathbf y)$ for convex $f$ and $g(\mathbf y) \le \mathcal L_g^{(r)}(\mathbf y)$ for concave $g$; and \emph{(ii)} $g(\mathbf y) \le \mathcal Q_g^{(r)}(\mathbf y) \triangleq \mathcal L_g^{(r)}(\mathbf y) + \tfrac{L}{2}\|\mathbf y - \mathbf y^{(r)}\|^2$ for $L$-smooth $g$~\cite{kannan13stochastic, blockage2025}. All bounds are global and tight at $\mathbf y^{(r)}$, so replacing each nonconvex term by its surrogate yields a convex restriction with monotonically convergent MM iterates~\cite{hu2025}.
\end{lemma}
\vspace{-0.5em}
 
By Lemma~\ref{lem:mm}(i), linearizing the functions $(T_{\rm tx}+\rho_k)^2$, $\log_2(\sum_{j\ne k}{\rm Tr}(\mathbf B_{\ell,k}\mathbf Q_j)+T_{\rm tx}\sigma_k^2)$, and $\|\textbf{Q}_k\|_2$ at the $r_1$-th iterate yields the affine surrogates $\mathcal{L}_{{\rm Obj},k}^{(r_1)}$, $\mathcal{L}_{{\rm C6a},\ell,k}^{(r_1)}$, and $\overline{\textbf{Q}}^{(r_1)}_k$, respectively; e.g., the spectral-norm subgradient gives $\overline{\textbf{Q}}^{(r_1)}_k = {\rm Tr}[\bm{\lambda}_{\rm max}(\textbf{Q}_k^{(r_1)})\bm{\lambda}_{\rm max}^H(\textbf{Q}_k^{(r_1)})(\textbf{Q}_k - \textbf{Q}_k^{(r_1)})] + \|\textbf{Q}_k^{(r_1)}\|_2$. By Lemma~\ref{lem:mm}(ii), the sigmoid $\phi_{\tau,\kappa}$, being $L_\phi$-smooth with $L_\phi = 0.1/(\tau^2{\rm sig}(\kappa))$~\cite{kannan13stochastic}, admits the quadratic majorizer $\phi_{\tau,\kappa}^{(r_1)}(\varepsilon_{\ell,k})$. Consequently, $(\widetilde{\widetilde{\rm C6a}}) \Rightarrow (\widetilde{\rm C6a})$, $(\widetilde{\widetilde{\rm C6b}}) \Rightarrow (\widetilde{\rm C6b})$, and $(\widetilde{\widetilde{\rm C9}}) \Rightarrow (\widetilde{\rm C9}) \Leftrightarrow ({\rm C9})$, while $\mathcal{L}_{{\rm Obj},k}^{(r_1)}$ lower-bounds the objective, such that a suboptimal solution to~\eqref{eq: formulation12} is obtained by solving, in the $(r_1+1)$-th iteration, as
\begin{align}\label{eq: formulation13}
&\underset{\mathcal{Q}, \ \bm{\rho}, \ T_{\rm pos}, \ T_{\rm tx}, \ \bm{\varepsilon}} {\rm{maximize}} \hspace{2mm} \sum\nolimits_{k\in\mathcal K}\tfrac{ (1-\delta_k)}{4 T_{\rm tot}} \left[ \mathcal{L}_{{\rm Obj},k}^{(r_1)} - (T_{\rm tx}-\rho_k)^2\right] \nonumber \\
&\rm{s.t.}\ ({\rm C2}), (\widetilde{\rm C3}), (\widetilde{\rm C4}), ({\rm C5}), (\widetilde{\rm C7}), ({\rm C8}),  \nonumber \\
&(\widetilde{\widetilde{\rm C6a}})\hspace{-1.3mm}:\hspace{-1mm} \log_2 \hspace{-1mm} \Big(\hspace{-1mm}\sum_{i \in \mathcal{K}}\hspace{-0.4mm} {\rm Tr}(\hspace{-0.3mm}\mathbf B_{\ell,k}\hspace{-0.5mm}\mathbf Q_i\hspace{-0.5mm}) \hspace{-1mm}+\hspace{-1mm} T_{\rm tx}\sigma_k^2 \hspace{-0.8mm}\Big) \hspace{-1mm}-\hspace{-0.8mm} \mathcal{L}_{{\rm C6a},\ell, k}^{(r_1)} \hspace{-1.5mm}- \hspace{-1mm}\rho_k \hspace{-1mm}\ge\hspace{-1mm} \varepsilon_{\hspace{-0.4mm} \ell,k}, \hspace{-0.9mm} \forall \ell \hspace{-0.2mm}, \hspace{-0.5mm}k\hspace{-0.2mm}, \nonumber \\
&(\widetilde{\widetilde{\rm C6b}})\hspace{-1mm}:\hspace{-3mm} \sum_{\ell\in\mathcal M_k} \hspace{-2.2mm}p_{\hspace{-0.3mm}\ell\hspace{-0.2mm},\hspace{-0.1mm}k} \hspace{-0.3mm} \phi_{\hspace{-0.6mm}\tau\hspace{-0.3mm},\kappa}^{(\hspace{-0.6mm}r_1\hspace{-0.6mm})}\hspace{-0.6mm}(\hspace{-0.7mm}\varepsilon_{\hspace{-0.4mm}\ell,k}\hspace{-0.6mm}) \hspace{-1.2mm}\le\hspace{-1.2mm} \delta_{\hspace{-0.3mm}k}\hspace{-0.3mm}, \hspace{-1mm}\forall \hspace{-0.3mm}k, (\widetilde{\widetilde{\rm C9}})\hspace{-1mm}:\hspace{-1mm} ||\textbf{Q}_k||_{\hspace{-0.3mm}*} \hspace{-1.3mm}-\hspace{-0.7mm} \overline{\textbf{Q}}_k^{(\hspace{-0.4mm}r_1\hspace{-0.4mm})} \hspace{-1.8mm}\le \hspace{-1.2mm}0\hspace{-0.25mm}, \hspace{-0.8mm}\forall k, 
\end{align}
which is convex and solvable by a standard convex programming solver. Since ${\rm Rank}(\textbf{Q}_k)\le 1$ is thereby guaranteed, both $\textbf{W}_k$ and $\textbf{w}_k$ are recovered from $\textbf{Q}_k = T_{\rm tx}\textbf{W}_k$.

\subsection{Solution of $\{\bm{\rho}, \mathbf{x}^{\rm Pin}, \bm{\varepsilon}, T_{\rm pos}, T_{\rm tx}\}$}
\label{subsec:subprob2}

Subproblem 2 is nonconvex for given $\mathcal{W}$, as the PA positions couple nonlinearly with the atom rates and LoS-state probabilities, while $T_{\rm tx}$ is bilinearly coupled with $\bm{\rho}$ in the objective. To decouple the position-induced amplitude and phase variations, we define $\textbf{C}_{\ell, k} = \text{diag} (\sqrt{\eta} \textbf{D}_{\ell} \textbf{w}_k^*)$, the element-wise positive amplitude vector $\textbf{v}_{k}$ with $[\textbf{v}_{k}]_n = e^{-\alpha (x_n^{\rm Pin} - x_{\rm B})}/\|\bm{\phi}_k - \bm{\psi}^{\rm{Pin}}_{n}\|$, and the unit-modulus phase vector $\textbf{a}_{k}$ with $[\textbf{a}_{k}]_n = e^{-j \nu_{n,k}}$, $\nu_{n, k} = \frac{2\pi}{\lambda}\| \bm{\phi}_k - \bm{\psi}^{\rm{Pin}}_{n}\| + \frac{2\pi}{\lambda_g}(x_n^{\rm Pin} - x_{\rm B})$, $\forall \ell, n, k$, and let $\varsigma_{n,k} \triangleq e^{-2\alpha (x_n^{\rm Pin} - x_{\rm B})}/[\textbf{V}_k]_{n,n}$ and $\mathcal K_{-k} \triangleq \mathcal K \backslash \{k\}$. Defining the auxiliary rank-one matrices $\textbf{A}_k = \textbf{a}_k \textbf{a}_k^H$ and $\textbf{V}_k = \textbf{v}_k\textbf{v}_k^H$, we have $\text{Tr}(\textbf{B}_{\ell, k} \textbf{W}_j) = \text{Tr}(\textbf{A}_k \textbf{C}_{\ell, j} \textbf{V}_k \textbf{C}^H_{\ell, j})$, and the Frobenius-norm identity gives the DC expression $\Gamma_{\ell,k}(\mathcal J) \triangleq \tfrac{1}{2}\sum_{i \in \mathcal{J}} [ \| \textbf{V}_{k} + \textbf{C}_{\ell, i}^{H} \textbf{A}_k \textbf{C}_{\ell, i} \|_{F}^2 - \mathcal{S}_{\ell,i,k} ] + \sigma_k^2$ with $\mathcal{S}_{\ell,i,k} \triangleq \|\textbf{V}_{k}\|_{F}^2 + \|\textbf{C}_{\ell, i}^{H} \textbf{A}_k \textbf{C}_{\ell, i}\|_{F}^2$. Introducing the slacks $\bm{\zeta} \hspace{-1mm}\triangleq\hspace{-1mm} \{\zeta_{\ell, k}^{\rm C6ab}, \zeta_{\ell, k}^{\rm C6ac} \hspace{-1mm}>\hspace{-1mm} 0\}$ and collecting the variables in $\mathcal{Z} \triangleq \{\bm{\rho}, \textbf{x}^{\rm Pin}, \bm{\varepsilon}, \bm{\zeta}, \bm{\nu}, \mathcal{A}, \mathcal{V}, T_{\rm pos}, T_{\rm tx}\}$, we recast subproblem 2 as
\begin{align}
&\underset{\mathcal{Z}} {\rm{maximize}} \hspace{2mm} \sum \nolimits _{k\in\mathcal K}\tfrac{ (1-\delta_k)}{4 T_{\rm tot}} \left[ (T_{\rm tx}+\rho_k)^2 - (T_{\rm tx}-\rho_k)^2\right]  \nonumber \\
 &\rm{s.t.}\ ({\rm C1}), ({\rm C2}), ({\rm C4}), ({\rm C5}), (\widetilde{\rm C7}), \nonumber \\
&(\widetilde{\overline{\rm C6aa}})\hspace{-1mm}:\hspace{-0.5mm} \log_2 (\zeta_{\ell, k}^{\rm C6ab} ) -\log_2 (\zeta_{\ell, k}^{\rm C6ac}) - \rho_k \ge \varepsilon_{\ell,k}, \forall \ell, k, \nonumber \\
&({\rm C6ab})\hspace{-1mm}:\hspace{-0.7mm} \zeta_{\ell, k}^{\rm C6ab} \hspace{-1.3mm}\le\hspace{-1mm} \Gamma_{\hspace{-0.8mm}\ell,k}\hspace{-0.4mm}(\hspace{-0.3mm}\mathcal K\hspace{-0.3mm})\hspace{-0.3mm}, \hspace{-0.6mm}\forall \ell,\hspace{-0.6mm}k, ({\rm C6ac})\hspace{-1mm}:\hspace{-0.7mm} \zeta_{\ell, k}^{\rm C6ac} \hspace{-1mm}\ge\hspace{-1mm} \Gamma_{\hspace{-0.8mm} \ell,k}\hspace{-0.3mm}(\hspace{-0.3mm}\mathcal K_{-k} \hspace{-0.4mm})\hspace{-0.3mm}, \hspace{-0.6mm} \forall \ell\hspace{-0.2mm}, \hspace{-0.5mm}k,\nonumber \\
&(\widetilde{\rm C6b})\hspace{-1mm}:\hspace{-0.5mm} \sum\nolimits_{\ell\in\mathcal M_k} p_{\ell,k}(\mathbf x^{\rm Pin}) \phi_{\tau,\kappa}\!\left(\varepsilon_{\ell, k}\right) \le \delta_k, \ \forall k, \nonumber \\
&({\rm C8a}),({\rm C8b})\hspace{-0.5mm}:  (x_n^{\rm Pin} - x_k)^2 + y_k^2 + z_n^2 \lesseqgtr \varsigma_{n,k}, \forall n,k, \nonumber \\
&({\rm C9a}),({\rm C9b})\hspace{-0.5mm}: \nu_{n,k} \lesseqgtr \tfrac{2 \pi}{\lambda} \sqrt{\varsigma_{n,k}} + \tfrac{2 \pi}{\lambda_g} (x_n^{\rm Pin} - x_{\rm B}), \forall n,k,\nonumber \\
&({\rm C10a})\hspace{-0.5mm}: \hspace{-0.5mm} \Re([\textbf{A}_k]_{\bar{n},1}) \hspace{-0.5mm}=\hspace{-0.5mm} \cos(\hat{\nu}_{\bar{n},k}), \forall \bar{n},k, \nonumber \\
&({\rm C10b})\hspace{-0.5mm}:\hspace{-0.5mm} \Im([\textbf{A}_k]_{\bar{n},1}) \hspace{-0.5mm}=\hspace{-0.5mm} -\sin(\hat{\nu}_{\bar{n},k}), \forall \bar{n},k, \nonumber \\
&({\rm C11})\hspace{-0.5mm}:\hspace{-0.5mm} \textbf{V}_k, \textbf{A}_k \bm{\succeq 0}, \forall k, \ ({\rm C12})\hspace{-0.5mm}:\hspace{-0.5mm} \text{Diag}(\textbf{A}_k) = \bm{1}_N,  \forall k, \nonumber \\
&({\rm C13}): \text{Rank}(\textbf{V}_k) = \text{Rank}(\textbf{A}_k) = 1, \forall k,
\label{eq: formulation2N}
\end{align}
where $\bar{n} = 2,\ldots,N$, $\hat{\nu}_{\bar{n},k} = \nu_{\bar{n},k} - \nu_{1,k}$, and $\lesseqgtr$ denotes the paired inequalities jointly enforcing the corresponding position--amplitude and position--phase equalities. Problem~\eqref{eq: formulation2N} remains nonconvex, owing to the DC objective and constraints $(\widetilde{\overline{\rm C6aa}})$--$({\rm C9a})$, the position-coupled $(\widetilde{\rm C6b})$, the periodic equalities $({\rm C10a})$, $({\rm C10b})$, and $({\rm C13})$.

To handle $(\widetilde{\rm C6b})$, we introduce slacks $\theta_{n,k} \hspace{-0.5mm}\in\hspace{-0.5mm} (0, 90^\circ]$, $l_{n,k} \hspace{-0.5mm}>\hspace{-0.5mm} 0$, and $\bm{\omega} \triangleq \{ \omega_{n,k}^{\rm C14a} < 0, \omega_{n,k}^{\rm C14b} \hspace{-0.5mm}\le\hspace{-0.5mm} 0, \omega_{n,k}^{\rm C14c} \le 0\}$, then
\begin{align}\label{eq: C6bb}
&(\widetilde{\overline{\rm C6b}})\hspace{-1mm}:\hspace{-2.8mm} \sum_{\ell\in\mathcal M_k} \hspace{-2.2mm} e^{\sum_{n \in \mathcal{N}}[b_{\ell, n} \omega_{n,k}^{\rm C14b} + (1-b_{\ell,n}) \omega_{n,k}^{\rm C14c}]} \phi_{\tau, \kappa}(\varepsilon_{\ell,k}) \hspace{-1mm}\le \hspace{-1mm}\delta_{k}, \forall k, \nonumber \\
&({\rm C14a}),({\rm C14b})\hspace{-1mm}:\hspace{-0.5mm} \omega_{n,k}^{\rm C14a} \le \mathcal{S}^{\rm C14b}_{n,k} \le \omega_{n,k}^{\rm C14b}, \forall n,k, \nonumber \\
&({\rm C14c})\hspace{-1mm}:\hspace{-0.5mm} \ln (1-e^{\omega_{n,k}^{\rm C14a}}) \le \omega_{n,k}^{\rm C14c},\forall n,k, \nonumber \\
&({\rm C15a}),({\rm C15b})\hspace{-1mm}:\hspace{-0.5mm} (x_n^{\rm Pin} - x_k)^2 + y_k^2 \lesseqgtr l_{n,k}^{2}, \forall n,k,\nonumber \\
&({\rm C16a}),({\rm C16b})\hspace{-1mm}:\hspace{-0.5mm} \tfrac{180}{\pi} \tan^{-1}(\tfrac{z_n}{l_{n,k}}) \lesseqgtr \theta_{n,k}, \ \forall n,k,
\end{align}
where $\mathcal{S}^{\rm C14b}_{n,k} \triangleq -\ln(1+A_L e^{-B_L(\theta_{n,k}-A_L)}) - \wp_L(l_{n,k}^2+z_n^2)$, $(\widetilde{\overline{\rm C6b}}) \Rightarrow (\widetilde{\rm C6b})$, and $({\rm C14b})$, $({\rm C14c})$, $({\rm C15b})$, $({\rm C16b})$ remain nonconvex but admit DC forms. Thus, applying Lemma~1 to $({\rm C13})$ yields the DC reformulation $(\overline{\rm C13})$, whose convex restriction $(\overline{\overline{\rm C13}}) \Rightarrow (\overline{\rm C13})$ follows by linearizing $\|\textbf{A}_k\|_2$ and $\|\textbf{V}_k\|_2$ as for $\|\textbf{Q}_k\|_2$ via Lemma~2(i).
 
At the $r_2$-th MM iterate, Lemma~2(i) is applied to all DC terms above: the bilinear objective is recast as in subproblem~1, whose convex $(T_{\rm tx}+\rho_k)^2$ admits the affine minorant $\mathcal{L}^{(r_2)}_{{\rm Obj},k}$, while the DC terms $\log_2(\zeta_{\ell,k}^{\rm C6ac})$, $\|\textbf{V}_k + \textbf{C}_{\ell,i}^H \textbf{A}_k \textbf{C}_{\ell,i}\|_F^2$, $\mathcal{S}_{\ell,j,k}$, $e^{-2\alpha(x_n^{\rm Pin}-x_{\rm B})}/[\textbf{V}_k]_{n,n}$, $(x_n^{\rm Pin}-x_k)^2$, $\tfrac{2\pi}{\lambda}\sqrt{e^{-2\alpha(x_n^{\rm Pin}-x_{\rm B})}/[\textbf{V}_k]_{n,n}}$, $\ln\phi_{\tau,\kappa}(\varepsilon_{\ell,k})$, $\mathcal{S}^{\rm C14b}_{n,k}$, $\ln(1-e^{\omega_{n,k}^{\rm C14a}})$, and $\tfrac{180}{\pi}\tan^{-1}(z_n/l_{n,k})$ yield the affine surrogates $\mathcal{L}^{(r_2)}_{{\rm C6aa},\ell,k}$, $\mathcal{L}^{(r_2)}_{{\rm C6ab},\ell,i,k}$, $\mathcal{L}^{(r_2)}_{{\rm C6ac},\ell,j,k}$, $\mathcal{L}^{(r_2)}_{{\rm C8a},n,k}$, $\mathcal{L}^{(r_2)}_{{\rm C8b},n,k}$, $\mathcal{L}^{(r_2)}_{{\rm C9a},n,k}$, $\mathcal{L}^{(r_2)}_{{\rm C6b},\ell,k}$, $\mathcal{L}^{(r_2)}_{{\rm C14b},n,k}$, $\mathcal{L}^{(r_2)}_{{\rm C14c},n,k}$, and $\mathcal{L}^{(r_2)}_{{\rm C16b},n,k}$, respectively. Consequently, $(\widetilde{\overline{\overline{\rm C6aa}}}) \Rightarrow (\widetilde{\overline{\rm C6aa}})$, $(\overline{\rm C6ab}) \Rightarrow ({\rm C6ab})$, $(\overline{\rm C6ac}) \Rightarrow ({\rm C6ac})$, $(\overline{\rm C8a})$--$(\overline{\rm C9a}) \Rightarrow ({\rm C8a})$--$({\rm C9a})$, $(\overline{\rm C14b})$--$(\overline{\rm C16b}) \Rightarrow ({\rm C14b})$--$({\rm C16b})$, $(\widetilde{\overline{\rm C6b}}) \Rightarrow \eqref{eq: C6bb}$, and $(\overline{\overline{\rm C13}}) \Rightarrow ({\rm C13})$. Finally, $({\rm C10a})$, $({\rm C10b})$ are handled by the penalty method~\cite{blockage2025}, augmenting the objective with $\Phi_{\bar{n},k} = (\Re([\textbf{A}_k]_{\bar{n},1}) - \cos\hat{\nu}_{\bar{n},k})^2 + (\Im([\textbf{A}_k]_{\bar{n},1}) + \sin\hat{\nu}_{\bar{n},k})^2$; as $\Phi_{\bar{n},k}$ is $L$-smooth with $L_{\rm AR} = L_{\rm AI} = 2$, $L_{\rm TH} = 4$~\cite{blockage2025}, Lemma~2(ii) gives the majorizer $\overline{\Phi}^{(r_2)}_{\bar{n},k} \ge \Phi_{\bar{n},k}$. Accordingly, the $(r_2+1)$-th iteration of the penalty-based MM method solves
\begin{align}
&\underset{\mathcal{Z}, \theta,\bm{\omega}}{\rm{maximize}} \hspace{0.5mm} \sum_{k\in\mathcal K}\hspace{-1mm}\tfrac{ (1-\delta_k)[ \mathcal{L}_{{\rm Obj},k}^{(r_2)} \hspace{-0.8mm}-\hspace{-0.3mm} (T_{\rm tx}\hspace{-0.4mm}-\hspace{-0.4mm}\rho_k)^2]}{4 T_{\rm tot}} - f_\text{P} \underset{k \in \mathcal{K}}{\sum} \overset{N}{\underset{\bar{n} = 2}{\sum}} \overline{\Phi}^{(\hspace{-0.3mm}r_2\hspace{-0.3mm})}_{\bar{n},k} \nonumber \\
\rm{s.t.}\ &({\rm C1}), ({\rm C2}), ({\rm C4}), ({\rm C5}), (\widetilde{\rm C7}), ({\rm C9b}), ({\rm C11}), ({\rm C12}), \nonumber \\
&(\overline{\overline{\rm C13}}), ({\rm C14a}), ({\rm C15a}), ({\rm C16a}), \nonumber \\
&(\widetilde{\overline{\overline{\rm C6aa}}})\hspace{-1.3mm}:\hspace{-0.8mm} \log_2 (\zeta_{\ell, k}^{\rm C6ab} ) -\mathcal{L}^{(r_2)}_{{\rm C6aa}, \ell, k} - \rho_k \ge \varepsilon_{\ell,k}, \forall \ell, k, \nonumber \\
&(\overline{\rm C6ab})\hspace{-1.3mm}:\hspace{-0.8mm} \zeta_{\ell, k}^{\rm C6ab} \le \tfrac{1}{2}\hspace{-1mm}\sum_{i \in \mathcal{K}} [ \mathcal{L}^{(r_2)}_{{\rm C6ab}, \ell, i, k} \hspace{-0.5mm}-\hspace{-0.5mm} \mathcal{S}_{\ell,i,k} ] + \sigma_k^2,  \forall \ell, k, \nonumber \\
&(\overline{\rm C6ac})\hspace{-1.3mm}:\hspace{-0.8mm} \zeta_{\ell, k}^{\rm C6ac} \hspace{-1mm}\ge \hspace{-1mm}\tfrac{\underset{j \ne k}{\sum}\hspace{-0.8mm}  \| \textbf{V}_{\hspace{-0.5mm}k} \hspace{-0.3mm}+ \textbf{C}_{\hspace{-0.3mm}\ell, j}^{H} \textbf{A}_k \textbf{C}_{\hspace{-0.3mm}\ell, j} \hspace{-0.3mm}\|_{\hspace{-0.3mm} F}^2 \hspace{-0.4mm}-\hspace{-0.3mm} \mathcal{L}^{(r_2)}_{{\rm C6ac}, \ell, j, k}}{2}\hspace{-0.8mm}+\hspace{-0.5mm} \sigma_k^2,  \forall \ell, k, \nonumber \\
&(\widetilde{\overline{\rm C6b}})\hspace{-1mm}:\hspace{-2.8mm} \sum_{\ell\in\mathcal M_k} \hspace{-2.2mm} e^{\underset{n \in \mathcal{N}}{\sum}[b_{\ell, n} \omega_{n,k}^{\rm C14b} + (1-b_{\ell,n}) \omega_{n,k}^{\rm C14c}] + \mathcal{L}^{(r_2)}_{{\rm C6b}, \ell, k}} \hspace{-1mm} \le \hspace{-1mm} \delta_{k}, \hspace{-1mm}\forall k,\nonumber \\
&(\overline{\rm C8a})\hspace{-1.3mm}:\hspace{-0.8mm} (x_n^{\rm Pin} - x_k)^2 + y_k^2 + z_n^2 \le \mathcal{L}^{(r_2)}_{{\rm C8a}, n, k}, \forall n,k, \nonumber \\
&(\overline{\rm C8b})\hspace{-1.3mm}:\hspace{-0.8mm} \mathcal{L}^{(r_2)}_{{\rm C8b}, n, k} + y_k^2 + z_n^2 \ge \tfrac{e^{-2\alpha (x_n^{\rm Pin} - x_{\rm B})}}{[\textbf{V}_k]_{n,n}}, \forall n,k, \nonumber \\
&(\overline{\rm C9a})\hspace{-1.3mm}:\hspace{-0.8mm} \mathcal{L}^{(r_2)}_{{\rm C9a}, n, k} \hspace{-1mm}\ge\hspace{-1mm} \nu_{n,k}, \hspace{-1mm}\forall \hspace{-0.3mm}n\hspace{-0.3mm},\hspace{-0.3mm}k\hspace{-0.3mm},  (\overline{\rm C14b})\hspace{-1.3mm}:\hspace{-0.8mm} \mathcal{L}^{(r_2)}_{\hspace{-0.3mm}{\rm C14b}\hspace{-0.3mm}, \hspace{-0.3mm}n\hspace{-0.3mm}, \hspace{-0.3mm}k} \hspace{-1.2mm}\le \hspace{-1mm}\omega_{n,k}^{\hspace{-0.4mm} {\rm C14b}}\hspace{-0.5mm}, \hspace{-1mm}\forall \hspace{-0.3mm}n\hspace{-0.3mm},\hspace{-0.3mm}k\hspace{-0.3mm}, \nonumber \\
&(\overline{\rm C14c})\hspace{-1.3mm}:\hspace{-1mm} \mathcal{L}^{(r_2)}_{\hspace{-0.3mm}{\rm C14c}\hspace{-0.15mm}, \hspace{-0.15mm}n\hspace{-0.15mm}, \hspace{-0.15mm}k} \hspace{-1.3mm}\le \hspace{-1.2mm} \omega_{n,k}^{\hspace{-0.55mm}{\rm C14c}}\hspace{-1mm}, \hspace{-0.6mm}\forall \hspace{-0.3mm}n\hspace{-0.3mm},\hspace{-0.3mm}k\hspace{-0.3mm},  (\overline{\rm C16b})\hspace{-1.3mm}:\hspace{-1mm} \mathcal{L}^{(r_2)}_{\hspace{-0.3mm}{\rm C16b}, n, k} \hspace{-1.3mm}\ge\hspace{-1.3mm} \theta_{\hspace{-0.5mm}n\hspace{-0.2mm},\hspace{-0.2mm}k}\hspace{-0.3mm},\hspace{-0.7mm}\forall \hspace{-0.3mm}n\hspace{-0.3mm},\hspace{-0.3mm}k\hspace{-0.3mm}, \nonumber \\
&(\overline{\rm C15b})\hspace{-1.3mm}:\hspace{-0.8mm} \mathcal{L}^{(r_2)}_{{\rm C8b}, n, k} + y_k^2 \ge l_{n,k}^{2}, \ \forall  n, k,
\label{eq: formulation2F}
\end{align}
where $f_{\rm P} \gg 1$ penalizes violations of $({\rm C10a})$, $({\rm C10b})$~\cite{zhao2026resource}. Problem~\eqref{eq: formulation2F} is convex and can be efficiently solved by standard convex solvers. Overall, the proposed BCD algorithm converges to a high-quality suboptimal solution to~\eqref{eq: formulation}.

\section{Simulation Results}
\begin{figure}[t]
	\centering
    \vspace{-0.8em}
	\subfigure[Performance versus $E^{\rm max}_{\rm total}$.]{
		\includegraphics[width=1.38in]{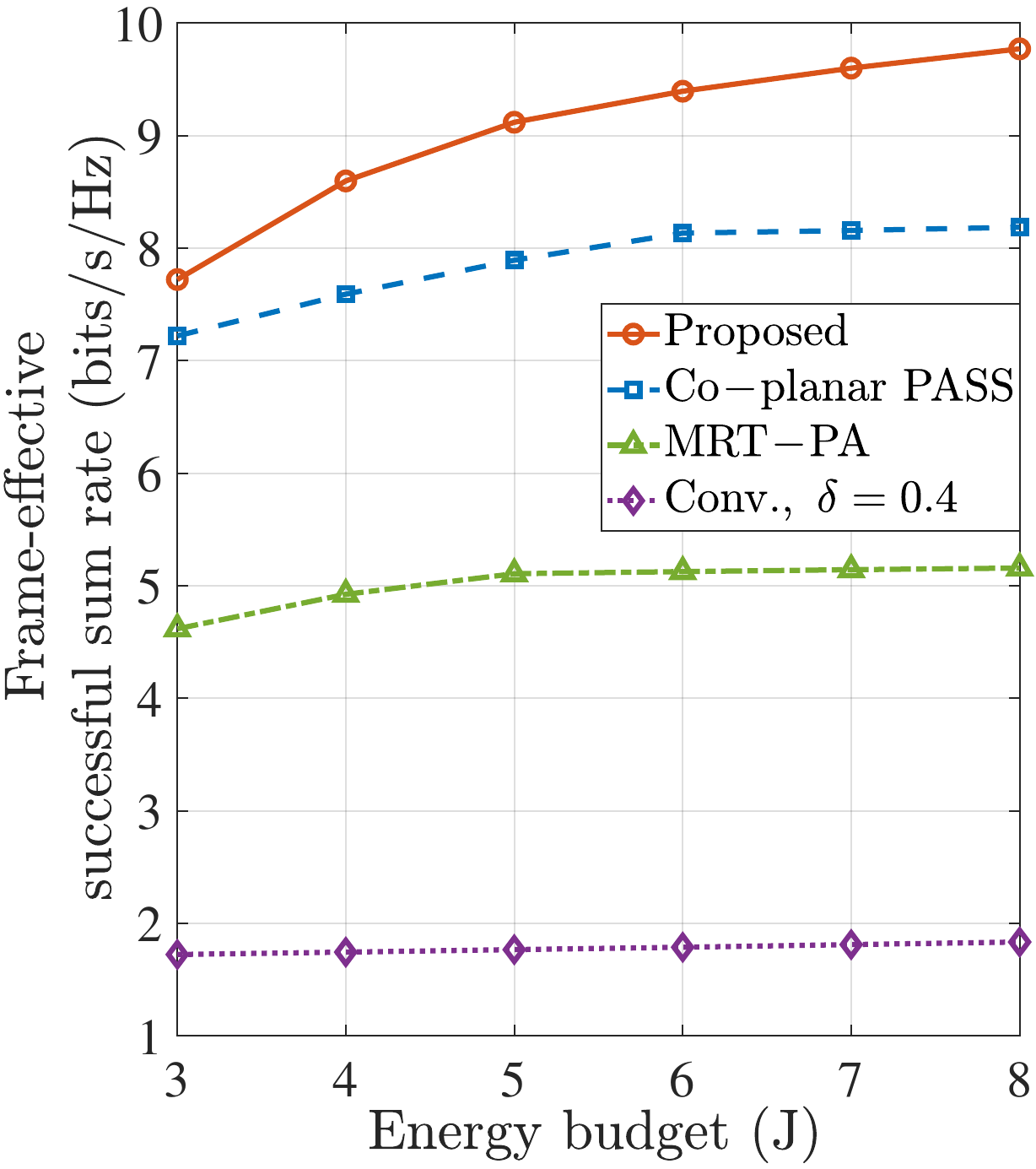}
	}\hspace{-0.52em}
	\subfigure[$K=3$ and $E^{\rm max}_{\rm total} = 6$ J.]{
		\includegraphics[width=1.675in]{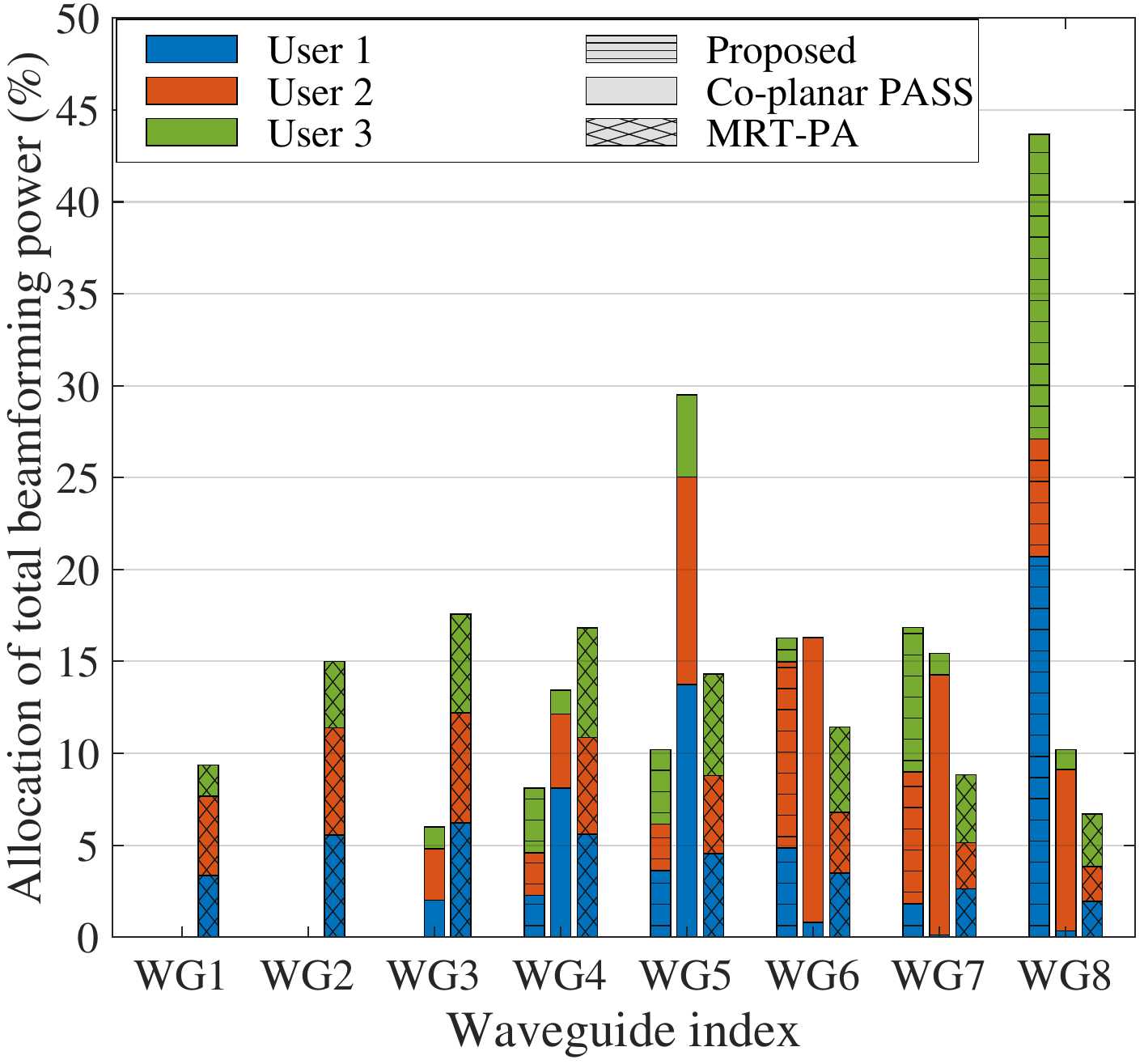}
	}
    \vspace{-0.8em}
	\caption{Average performance for different antenna settings, with a snapshot example of the per-waveguide beamforming power allocation.}
	\label{fig:energy}
    \vspace{-2em}
\end{figure}
We evaluate the proposed vertically layered PASS at $f_c=28$~GHz, where $K=3$ users are randomly located in a $15\times15$~m$^2$ area and the BS at $\bm{\psi}^{\rm BS}=[0,0,5]^\top$~m is connected to $N=8$ waveguides with $L=9$~m and $z_n=5+(n-1)$~m.  Unless stated otherwise, $\eta_{\rm eff}=1.42$, $\epsilon_r=2.1$, $\tan(\delta_e)=2\times10^{-4}$, $A_{\rm L}=11$, $B_{\rm L}=0.13$, $\wp_{\rm L}=2\times10^{-5}$~m$^{-2}$, $\kappa_{\rm f} = 1$~dB/km, $\sigma_k^2 = -100$~dBm,
$P^{\max}_{\rm Tx}=30$~dBm, $\eta_{\rm BS}=0.4$, $P_{\rm cir}=0.05$~W,
$T_{\rm tot}=2$~s, $P_{\rm mot}=0.1$~W, and
$v_{\rm PA}=1$~m/s~\cite{xu2025joint,li2026power}.
Three benchmarks are considered: \emph{Co-planar PASS} deploys all waveguides at $z=5$~m while retaining the proposed optimization framework; \emph{MRT-PA} retains the vertical deployment and PA positioning but adopts maximum ratio transmission with a fixed transmission time; and \emph{Conv.} employs a fixed antenna array, optimizing only the beamforming, coding rates, and transmission time.  We set $\delta_k=0.1$, $\forall k$, except for \emph{Conv.}, which is infeasible under this target and thus is evaluated at $\delta_k=0.4$ and assessed by the delivered sum rate in (5) without the $(1-\delta)$ weighting.

Fig.~\ref{fig:energy}(a) shows that the successfully delivered sum rate first increases with the energy budget and eventually saturates as the transmit power and frame duration constraints become dominant.  The \emph{Proposed} scheme consistently outperforms the benchmarks, since the vertical DoF enables more effective joint optimization over propagation loss, attenuation, repositioning cost, and spatial channel dynamics.  Moreover, \emph{MRT-PA} exhibits a larger gap than the co-planar PASS, indicating that the loss from a prescribed beamformer and a fixed transmission time outweighs the gain from vertical deployment, which confirms the necessity of jointly optimizing the beamforming and time allocation.  As expected, \emph{Conv.} performs worst despite its relaxed outage target, as a fixed array can neither adapt its geometry nor exploit aperture mobility. The allocation strategy underpinning these gains is illustrated in Fig.~\ref{fig:energy}(b), where the waveguides (WGs) are indexed by height for the proposed scheme and by horizontal location for the co-planar PASS. It is observed that the proposed design concentrates $43\%$ of the total power on WG8.
This allocation reflects the favorable tradeoff offered by the upper layer: the LoS availability increases, e.g., from $(0.95, 0.57, 0.81)$ on WG1 to $(0.99, 0.97, 0.99)$ on WG8 across $(u_1, u_2, u_3)$. Thus, the LoS reliability gain from exploiting WG8 outweighs the associated repositioning energy costs and increased attenuation distance.
In contrast, the co-planar PASS lacks height diversity and thus relies more heavily on horizontal PA repositioning, whereas \emph{MRT-PA} spreads its power across all WGs, as its prescribed beamformer cannot suppress the multiuser interference.


\begin{figure}[t]
\centerline{\includegraphics[width=3.2in]{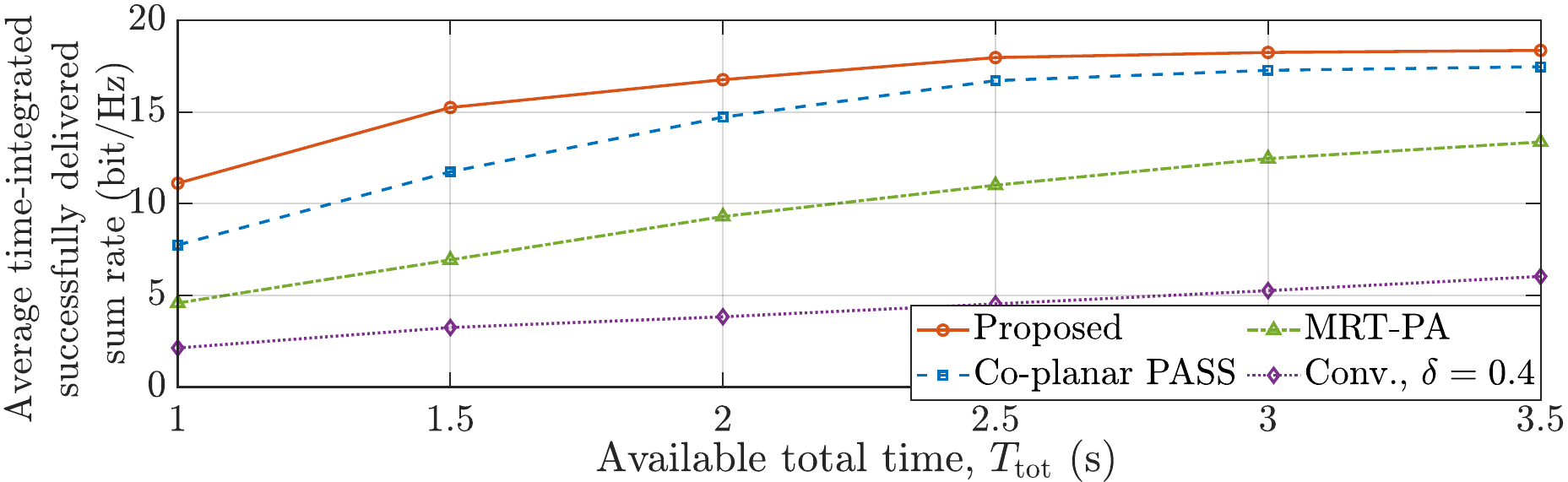}}
\vspace{-1em}
\caption{Average time-integrated successfully delivered sum rate versus the available total time $T_{\rm tot}$ for different antenna systems under $E^{\max}_{\rm total}=5$ J.}
\vspace{-2em}
\label{fig:TimeS}
\end{figure}

Since $T_{\rm tot}$ varies as an available-time budget, we report the time-integrated successfully delivered sum rate rather than the frame-normalized rate, avoiding an artificial normalization penalty for large deadlines.
Fig.~\ref{fig:TimeS} shows that the proposed scheme attains the highest sum rate for all considered $T_{\rm tot}$, with the largest gain in the short-deadline regime, as its height diversity provides LoS-favorable branches without relying heavily on horizontal repositioning, such that less positioning time is required and more of the frame is reserved for transmission. Lacking such diversity, the co-planar PASS depends more strongly on horizontal PA placement, such that additional time partially narrows the gap. In contrast, \emph{MRT-PA} exploits the additional time only for PA repositioning, and its fixed beamformer cannot accurately mitigate the multiuser interference, such that the gap persists across the entire range.
The fixed array remains the least effective and falls further behind. However, its sum rate keeps increasing as the additional time is devoted entirely to transmission without PA repositioning.

\section{Conclusion}

This paper investigated an energy-aware vertically layered PASS capturing in-waveguide attenuation, fronthaul loss, and repositioning energy under a hybrid elevation- and distance-dependent LoS availability model. The beamforming, PA positions, time allocation, and coding rates were jointly optimized to maximize the expected successfully delivered sum rate under probabilistic outage constraints. We derived a CM outage envelope and a differentiable safe approximation, and developed a penalty-aided MM-based BCD algorithm. Simulation results confirmed the vertical-layering gains and the reduced repositioning time over the co-planar PASS, and unveiled the fundamental energy–LoS tradeoff, despite LoS-favorable positions generally incurring higher path attenuation.


\end{document}